\documentclass[aps,twocolumn,floatfix,superscriptaddress,preprintnumbers,showpacs,showkeys,nofootinbib]{revtex4}

\usepackage{amssymb}
\usepackage{amsmath}
\usepackage{amsfonts}
\usepackage{graphicx}
\usepackage{color}
\usepackage{xspace}
\usepackage{ulem}
\usepackage{lscape}
\usepackage{ wasysym }
\usepackage[toc,page]{appendix}
\usepackage{cancel}

\usepackage{multirow,rotating}

\def\gsim{\raise0.3ex\hbox{$\;>$\kern-0.75em\raise-1.1ex\hbox{$\sim\;$}}}
\def\lsim{\raise0.3ex\hbox{$\;<$\kern-0.75em\raise-1.1ex\hbox{$\sim\;$}}}

\def\znbb{0\nu\beta\beta}

\newcommand{\ba}[1]{\begin{eqnarray} \label{(#1)}}
\newcommand{\ea}{\end{eqnarray}}

\newcommand{\AddrAHEP}{
  {\it AHEP Group, Instituto de F\'{\i}sica Corpuscular --
    C.S.I.C./Universitat de Val{\`e}ncia \\
    Edificio de Institutos de Paterna, Apartado 22085,
  E--46071 Val{\`e}ncia, Spain}}

\newcommand{\AddrUFSM}{
Universidad T\'ecnica Federico Santa Mar\'\i a, \\ 
Centro-Cient\'\i fico-Tecnol\'{o}gico de Valpara\'\i so, \\ 
Casilla 110-V, Valpara\'\i so,  Chile}

\def\gsim{\raise0.3ex\hbox{$\;>$\kern-0.75em\raise-1.1ex\hbox{$\sim\;$}}}
\def\lsim{\raise0.3ex\hbox{$\;<$\kern-0.75em\raise-1.1ex\hbox{$\sim\;$}}}

\begin{document}

\preprint{IFIC/17-50}  

\title{Neutrinoless double beta decay and QCD running at low energy
  scales}

\author{M. Gonz\'alez} \email{marcela.gonzalezp@usm.cl}\affiliation{\AddrUFSM}
\author{M. Hirsch} \email{mahirsch@ific.uv.es}\affiliation{\AddrAHEP}
\author{S.G. Kovalenko}\email{sergey.kovalenko@usm.cl}\affiliation{\AddrUFSM}



\begin{abstract}
There is a common belief that the main uncertainties in the
theoretical analysis of neutrinoless double beta ($\znbb$) decay
originate from the nuclear matrix elements. Here, we uncover another
previously overlooked source of potentially large uncertainties
stemming from non-perturbative QCD effects.  Recently perturbative QCD
corrections have been calculated for all dimension 6 and 9 effective
operators describing $\znbb$-decay and their importance for a reliable
treatment of $\znbb$-decay has been demonstrated.  However, these
perturbative results are valid at energy scales above $\sim 1$ GeV,
while the typical $0\nu\beta\beta$-scale is about $\sim 100$ MeV.  In
view of this fact we examine the possibility of extrapolating the
perturbative results towards sub-GeV non-perturbative scales on the
basis of the QCD coupling constant ``freezing'' behavior using
Background Perturbation Theory. Our analysis suggests that such an
infrared extrapolation does modify the perturbative results for both
short-range and long-range mechanisms of $\znbb$-decay in general only
moderately. We also discuss that the tensor$\otimes$tensor
  effective operator can not appear alone in the low-energy limit of
  any renormalizable high-scale model and then demonstrate that all
  five linearly independent combinations of the scalar and tensor
  operators, that can appear in renormalizable models, are infrared
  stable.

\end{abstract}

\maketitle


\section{Introduction}
\label{sec:introduction}
From neutrino oscillation experiments it is nowadays well-known that
at least two neutrinos have non-zero masses. Oscillation experiments,
however, can not decide whether neutrinos are Dirac or Majorana
particles.  Lepton number violating (LNV) processes have to be studied
to explore the neutrino nature
\cite{Schechter:1981bd,Duerr:2011zd,Hirsch:2006yk}. Neutrinoless
double beta decay ($\znbb$), for recent reviews see for instance
\cite{Engel:2016xgb,Deppisch:2012nb,Rodejohann:2011mu}, is by far the
most powerful available probe of lepton number violation, and its
non-observation allows to constrain LNV beyond standard model (BSM)
physics. From several experimental searches for $\znbb$
\cite{Albert:2014awa,KamLAND-Zen:2016pfg,Agostini:2017iyd,GERDA_Taup:2017}
the best current bound on the $\znbb$ half-life, $T_{1/2}^{0\nu}$, comes
from the KamLAND-Zen experiment \cite{KamLAND-Zen:2016pfg}:
\begin{eqnarray}
T^{0\nu}_{1/2}({}^{136}{\rm Xe}) & \ge & 1.07\times 10^{26} \ {\rm ys} 
\ (90\% {\rm C.L.}).
\label{eq:TXe} 
\end{eqnarray}

From the theoretical perspective, there are two different kind of
contributions to the $\znbb$ amplitude: the short-range mechanisms
(SRM) \cite{Pas:2000vn}, which are mediated by heavy particle
exchange; and the long-range mechanisms (LRM) \cite{Pas:1999fc}, in
which a light neutrino is exchanged between two point-like vertices.
It has been recently demonstrated that QCD corrections to $\znbb$ are
important, especially in the SRM case \cite{Gonzalez:2015ady} due the
presence of the color-mismatch effect and the corresponding mixing of
different operators, with numerically very different nuclear matrix
elements (NME). This effect leads to differences in the limits on the
Wilson Coefficients (WC), which amount in some cases up to 3 orders of
magnitude.  On the other hand, the LRM operate between two different
and distant nucleons, so that no color-mismatch appears and only QCD
vertex corrections have to be taken into account. Their effect on the
extracted limits does not exceed 60\% \cite{Arbelaez:2016zlt}, less
than the typical estimate of the uncertainties of the nuclear matrix
elements (NMEs).

These QCD RGE results \cite{Gonzalez:2015ady,Arbelaez:2016zlt} are
valid for energy scales above $\sim 1$ GeV - the limit of perturbative
QCD.  On the other hand, the typical scale of $\znbb$-decay is about
$\sim 100$ MeV. It is then natural to ask, if these perturbative
results still can be considered a reasonable approximation or whether
they could be drastically affected by some non-perturbative effects.
In this paper, we try to address this issue within the
  semi-empirical approach of "freezing" the running of the strong
  coupling constant at low energies. Freezing is motivated by some
  approaches to non-perturbative QCD effects, in particular, by the
  \textit{Background Perturbation Theory} (BPTh) \cite{Abbott:1980hw}.
 We do not pretend that our work will determine the final
  conclusion about this question, but rather hope that our analysis
will allow us at least to visualize this problem, overlooked in the
literature, and stimulate efforts for theoretically grounded studies
in this direction.  The conventional approach would rely on the
operator product expansion for observables, such as $\znbb$ half-life,
and on a proper matching of the quark and nucleon level theories at a
certain scale $\mu_{0}$. This is a low-energy scale, down to which
perturbative QCD for the quark-level theory is valid. In a
self-consistent approach both Wilson coefficients and the nucleon
matrix elements of the effective operators depend on $\mu_{0}$ in such
a way that  the effect in any observable should cancel, in
  theory. Unfortunately, the nucleon matrix elements of
$\znbb$-operators have not been studied yet from this perspective. The
hadronization prescriptions existing in the literature (see, for
instance, Refs.~\cite{Hirsch:1995ek,Faessler:1996ph}) parameterize the
matrix elements in terms of phenomenological nucleon form-factors
loosing connection with QCD and, therefore, a dependence on the
matching scale $\mu_{0}$  remains.  Presumably, lattice QCD is
able to shed light on this issue. 
Some recent lattice QCD publications about matrix elements of $\znbb$-operators can be found in Refs. \cite{Nicholson:2016byl,Nicholson:2018mwc, Cirigliano:2017ymo}.
Recently an approach to hadronization in
$\znbb$-decay, which probably suits better for the matching with
lattice QCD, has been developed in 
Refs.~\cite{Prezeau:2003xn,Cirigliano:2017tvr}
on
the basis of chiral effective field theory. However, the issue of the
matching-scale dependence still remains unaddressed.  This is the
situation in which we adopted the idea of ``freezing''. 
In a 
sense it can be thought of as a rough modeling of the matching
scale-dependence of the nucleon matrix elements alleviating the
dependence of the $\znbb$ half-life on this scale.

\section{QCD running coupling constant in the infrared limit}
\label{sec:QCD running coupling constant in the infrared limit}

The high-energy behavior of QCD is well understood. Thanks to
asymptotic freedom, perturbation theory allows an accurate calculation
of the running QCD coupling constant $\alpha_{s}(\mu)$.  In the
infrared region, however, non-perturbative effects become important
and lead, in particular, to color confinement at some energy scale
below $\sim 1$ GeV
\cite{Prosperi:2006hx,Altarelli:2013bpa,Deur:2016tte}.  At such low
energies, perturbative QCD (pQCD) suffers a singularity, the so-called
Landau pole, i.e. $\alpha_{s}(\mu)$ tends to infinity. This
singularity prohibits even a formal extrapolation of the pQCD results
to the region below $\sim 1$ GeV.  Unsurprisingly, in the literature
there is a plethora of approaches aiming at the extrapolation of the
pQCD results towards the infrared (IR) region taking into account at
least some non-perturbative effects suggested by QCD. We will not
discuss all of these attempts and instead refer to the recent review
\cite{Deur:2016tte} on this subject.

However, different authors have discussed the possibility, for a list
of references see for example \cite{Deur:2016tte}, that the
predictions of QCD can be formulated in terms of a modified
$\tilde\alpha_s(\mu)$, which is finite in the IR regime. Basically,
the strong increase of $\alpha_s$, predicted by pQCD, is thought to be
regularized by non-perturbative effects summed up in an effective
$\tilde\alpha_s(\mu)$, which stops growing at some infrared
scale. This IR behavior, dubbed ``freezing'' of $\alpha_s$ in the
literature is compatible with the lattice QCD simulations
\cite{Lattice Data,Aguilar}.

To give an example let us consider the \textit{Background Perturbation
  Theory} (BPTh) \cite{Abbott:1980hw} applied to QCD. In this approach
the gluon field $A_{\mu}$ is separated into two parts
$A_{\mu}=a_{\mu}+B_{\mu}$, the perturbative part $a_{\mu}$, and an
effective non-perturbative background field $B_{\mu}$.  The BPTh
one-loop QCD running coupling constant has been calculated in Refs.~
\cite{Simonov:1993kt,Simonov:2010gb}.  It has been shown that the
non-perturbative background field generates an effective mass $M_{B}$
for the gluon in such a way that the argument of $\alpha_s(\mu^2)$ is
replaced by $\mu^2\rightarrow\mu^2+M_{B}^2$. In the UV limit the
effects of the background field become negligible, while at low
energies it essentially leads to a cutoff in the running. Then, the
modified one-loop QCD running coupling constant can be written as:
\begin{equation}
  \tilde\alpha_s(\mu^2)=
  \frac{\alpha_s(\lambda)}{1+\beta_0\frac{\alpha_{s}(\lambda)}{4\pi} \log{\frac{\mu^2+M_{B}^2}{\lambda^2}}}
\label{eq:alphafr}
\end{equation}
Here, $\beta_0$ is the usual $\beta$-coefficient, while the mass
parameter, $M_{B}$, is non-universal and depends on the specific
process in consideration \cite{Badalian:1997de,Simonov:2010gb}.

The parameter $M_B$ can be related to the mass of the glueball
$M_{2g}(0^{++})$ \cite{Simonov:2010gb}, a bound-state of two gluons
connected with the adjoint string
$M=M_B=M_{2g}(0^{++})=\sqrt{2\pi\sigma_a}\sim 2$ GeV, or in the case
of \textit{static $Q\bar Q$ potential}
$M=M_{Q\bar{Q}}=\sqrt{2\pi\sigma_f}\sim 1$ GeV.  In
Refs.~\cite{Badalian:2000hv,Badalian:2001by} $\alpha_s$ has been
calculated in the BPTh up to 3 loops and the value $M_B=1.06$ GeV has
been extracted from the analysis of bottomium fine structure.  We
mention also \cite{Badalian:2004xv}, quoting $M_B=0.95$ GeV, and
\cite{Simonov:1993kt}, which give $M_B=2.15$ GeV.

Below we revisit the QCD corrections to the SRM and LRM of
$0\nu\beta\beta$-decay \cite{Gonzalez:2015ady,Arbelaez:2016zlt}
applying Eq.~(\ref{eq:alphafr}) in order to extrapolate the pQCD
results towards $\mu\sim 100$~MeV, the characteristic scale of
$0\nu\beta\beta$-decay. In order to estimate the uncertainty of our
results, we will not use some specific value of $M_B$ and instead show
our results as functions of the calculated value of the ``frozen''
value of $\alpha_S^F$, corresponding to $\tilde\alpha_s(\mu^2)$ for
$\mu\le 0.1$ GeV. Although not numerically relevant, for completeness
we mention that for the normalization of the expression in
Eq.~(\ref{eq:alphafr}), we use the experimental value
$\alpha_s(\mu=M_Z)$ from \cite{Beringer:1900zz}.

\section{QCD running of Wilson coefficients}

Here we apply the IR extrapolation based on ``freezing'', discussed
above, to the pQCD results for $\znbb$-decay derived in refs
\cite{Gonzalez:2015ady,Arbelaez:2016zlt,Arbelaez:2016uto}.

\subsection{pQCD Wilson Coefficients}

First let us recall the pQCD results for the short-range and
long-range mechanisms of $\znbb$-decay derived in
\cite{Gonzalez:2015ady,Arbelaez:2016uto}.

{\bf (A)} {\it Short-range mechanisms} of $\znbb$-decay are mediated
by the exchange of heavy particles of mass $M_{I}$ between two
nucleons of the decaying nucleus.  At low energy scales $\mu<M_I$ this
mechanism is described the low-energy effective Lagrangian
\cite{Pas:2000vn,Gonzalez:2015ady}:
\begin{eqnarray}\label{eq:LagGen-Short}
{\cal L}^{0\nu\beta\beta}_{\rm eff} = \frac{G_F^2}{2 m_p} \,
              \sum_{i, XY} C_{i}^{XY}(\mu)\cdot \mathcal{O}^{(9)XY}_{i}(\mu),
\end{eqnarray}
with the complete set of dimension-9 non-equivalent $\znbb$-operators
\cite{Gonzalez:2015ady,Arbelaez:2016uto}
\begin{eqnarray}
\label{eq:OperBasis-1}
\mathcal{O}^{(9)XY}_{1}&=& 4 ({\bar u}P_{X}d) ({\bar u}P_{Y}d) \ j,\\
\label{eq:OperBasis-2}
\mathcal{O}^{(9)XX}_{2}&=& 4 ({\bar u}\sigma^{\mu\nu}P_{X}d)
                         ({\bar u}\sigma_{\mu\nu}P_{X}d) \ j,\\
\label{eq:OperBasis-3}
\mathcal{O}^{(9)XY}_{3}&=& 4 ({\bar u}\gamma^{\mu}P_{X}d) 
                        ({\bar u}\gamma_{\mu}P_{Y}d) \  j,\\
\label{eq:OperBasis-4}
\mathcal{O}^{(9)XY}_{4}&=& 4 ({\bar u}\gamma^{\mu}P_{X}d) 
                         ({\bar u}\sigma_{\mu\nu}P_{Y}d) \ j^{\nu},\\
\label{eq:OperBasis-5}
\mathcal{O}^{(9)XY}_{5}&=& 4 ({\bar u}\gamma^{\mu}P_{X}d) ({\bar u}P_{Y}d) \ j_{\mu},
\end{eqnarray}
where $X,Y = L,R$ and the leptonic currents are 
\begin{eqnarray}\label{eq:Curr}
j = {\bar e}(1\pm \gamma_{5})e^c \, , \quad j_{\mu} = {\bar e}\gamma_{\mu}\gamma_{5} e^c .
\end{eqnarray}
The QCD corrections induce operator mixing due to the color-mismatch
effect, which is equivalent to mixing different Wilson coefficients
$C_{i}^{XY}(\mu)$ in Eq.~(\ref{eq:LagGen-Short}) in the
amplitude. This effect has a dramatical impact on the numerical
predictions of high-scale models for the $\znbb$-decay half-life
\cite{Gonzalez:2015ady,Arbelaez:2016uto} due to the significant
differences in the numerical values of NMEs of different operators
(\ref{eq:OperBasis-1})-(\ref{eq:OperBasis-5}).  The QCD-corrected
$\znbb$-decay half-life formula, reads \cite{Gonzalez:2015ady}
\begin{eqnarray}\label{eq:T12}  
\Big[ T^{\znbb}_{1/2}\Big]^{-1} &=& 
G_1 \left|\sum_{i=1}^{3} \beta_{i}^{(SRM)XY}(\mu_{0}, \Lambda) C^{XY}_{i}(\Lambda) \right|^2 \\
&& +  
G_2 \left|\sum_{i=4}^{5} \beta_{i}^{(SRM)XY}(\mu_{0}, \Lambda) C^{XY}_{i}(\Lambda)\right|^2 \nonumber
\end{eqnarray}
$G_{1,2}$ are phase space factors \cite{Doi:1985dx,Pas:2000vn}, and
summation over the different chiralities $X,Y = L,R$ is implied.  The
parameters $\beta_{i}^{(SRM)}$ are defined in \cite{Gonzalez:2015ady}
and represent linear combinations of the NMEs of the operators
(\ref{eq:OperBasis-1})-(\ref{eq:OperBasis-5}).

{\bf (B)} {\it Long-Range Mechanisms} describe contributions via light
neutrino exchange between two quarks belonging to two different and
distant nucleons of a decaying nucleus. One of the vertices connected
by the neutrino propagator is the standard $V-A$ vertex while the
other is an LNV $\Delta L =2$ beyond the SM (BSM) vertex originating
from some heavy particle exchange. At low energies the BSM vertices
are given in terms of the effective operators \cite{Arbelaez:2016uto}
\begin{eqnarray}
\label{eq:hc3-1}
\mathcal{O}_{1}^{(6) X} &=& 4 (\bar{u} P_{X} d)  \left(\bar{e} P_{R} \nu^{C}\right), \\ 
\label{eq:hc3-2}
\mathcal{O}_{2}^{(6) X} &=& 4  (\bar{u} \sigma^{\mu\nu}P_{X} d) \left(\bar{e}\sigma^{\mu\nu}  P_{R} \nu^{C}\right),\\
\label{eq:hc3-3}
\mathcal{O}_{3}^{(6) X} &=& 4  (\bar{u} \gamma_{\mu}P_{X} d) 
\left(\bar{e} \gamma^{\mu} P_{R} \nu^{C}\right)
\end{eqnarray}
with $X= R, L$.  Then $\znbb$-decay amplitude is given by the
second-order of perturbation theory in the effective Lagrangian
\cite{Pas:1999fc,Arbelaez:2016uto}:
\begin{equation}
\mathcal{L}_{\rm eff}^{d=6}=\frac{G_F}{\sqrt{2}} \left( j^{\mu}J^{\dag}_{\mu}+  
\sum_{i}C^{X}_{i}(\mu)\mathcal{O}^{(6)X}_{i}(\mu)\right). 
\label{eq:lrlagrangian}
\end{equation}
Here the first term is the SM low-energy 4-fermion effective
interaction of the currents
\begin{eqnarray}\label{eq:V-A}
&&j^{\mu} = \bar{e}\gamma^{\mu}(1 - \gamma_{5}) \nu, \ \ \ 
J_{\mu} = \bar{d}\gamma_{\mu}(1 - \gamma_{5}) u.
\end{eqnarray} 
The QCD-corrected $0\nu\beta\beta$-decay half-life formula  for the LRM reads 
\cite{Arbelaez:2016zlt} 
\begin{eqnarray}\label{eq:T-QCD-LRM} 
\left[T^{\znbb}_{1/2}\right]^{-1}= \sum\limits_{i, X} G_{0i}\left|\beta^{(LRM)X}_{i}(\mu_{0}, \Lambda) C^{X}_{i}(\Lambda)\right|^2,
\end{eqnarray}
where $G_{0i}$ and $({\rm NME})_{i}$ are the phase-space factors
\cite{DKT}. The coefficients $\beta^{(LRM)}_{i}(\mu_{0}, \Lambda)$
involve NMEs of the operators (\ref{eq:hc3-1})-(\ref{eq:hc3-3}) and
the pQCD RGE running parameters.  Unlike the analogous
$\beta$-coefficients in Eq.~(\ref{eq:T12}) they do not mix the NMEs of
different operators. This is a drastic difference between the SRM and
LRM caused by the absence of the color-mismatch in the latter case as
demonstrated in Ref.~\cite{Arbelaez:2016uto}.

\subsection{``Freezing'' Wilson Coefficients}

The Wilson coefficients $C_i$ in Eqs.~(\ref{eq:LagGen-Short}) and
(\ref{eq:lrlagrangian}) are calculable in the framework of a
particular underlying model above some high-energy scale $\Lambda$.
These coefficients can be expressed in terms of the model couplings
and heavy particle masses (for a comprehensive analysis see
Ref.~\cite{Arbelaez:2016uto}).  At the scale $\Lambda$ the model is
matched to the effective theory given by the Lagrangians
(\ref{eq:LagGen-Short}) and (\ref{eq:lrlagrangian}). Then the Wilson
coefficients $C_i$ should be RGE evolved, taking into account the QCD
loop-corrections, down to the characteristic scale $\mu_{0}$ of
$\znbb$, typically about 100~MeV.
Refs.~\cite{Gonzalez:2015ady,Arbelaez:2016zlt} stopped the RGE
evolution at 1~GeV. This kind of truncation is a common practice in
the literature applying pQCD to observable processes. We now will
discuss the crucial question if the pQCD results are stable on the way
from 1~GeV to 100~MeV.

The discussion in section (\ref{sec:QCD running coupling constant in the
infrared limit}) suggests a method for a rough assessment of the
effect of the IR extrapolation by simple freezing the one-loop pQCD
running coupling. This amounts to the replacement of $\alpha_{s}(\mu)$
by $\tilde\alpha_{s}(\mu)$ from Eq.~(\ref{eq:alphafr}) in the
expressions for all the coefficients $\beta (\mu_{0}, \Lambda)$ in
Eqs.~(\ref{eq:T12}) , (\ref{eq:T-QCD-LRM}). Since
$\tilde\alpha_{s}(\mu)$ is finite in the IR limit,
one can then extend the running down to the required value of
$\mu_0 \simeq$ 100~MeV.

\begin{figure}[t]

\includegraphics[width=1.0\linewidth]{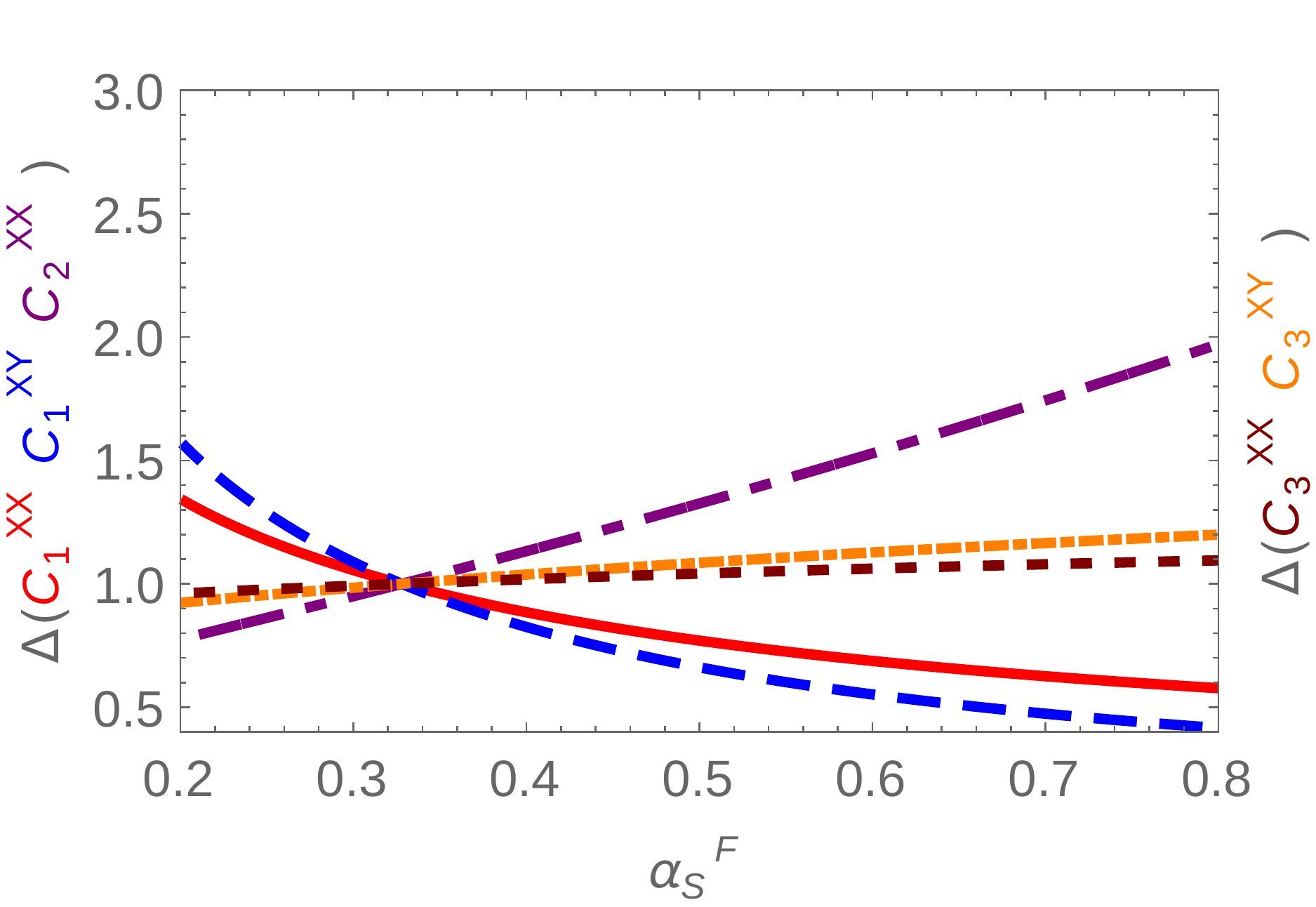}
\includegraphics[width=1.0\linewidth]{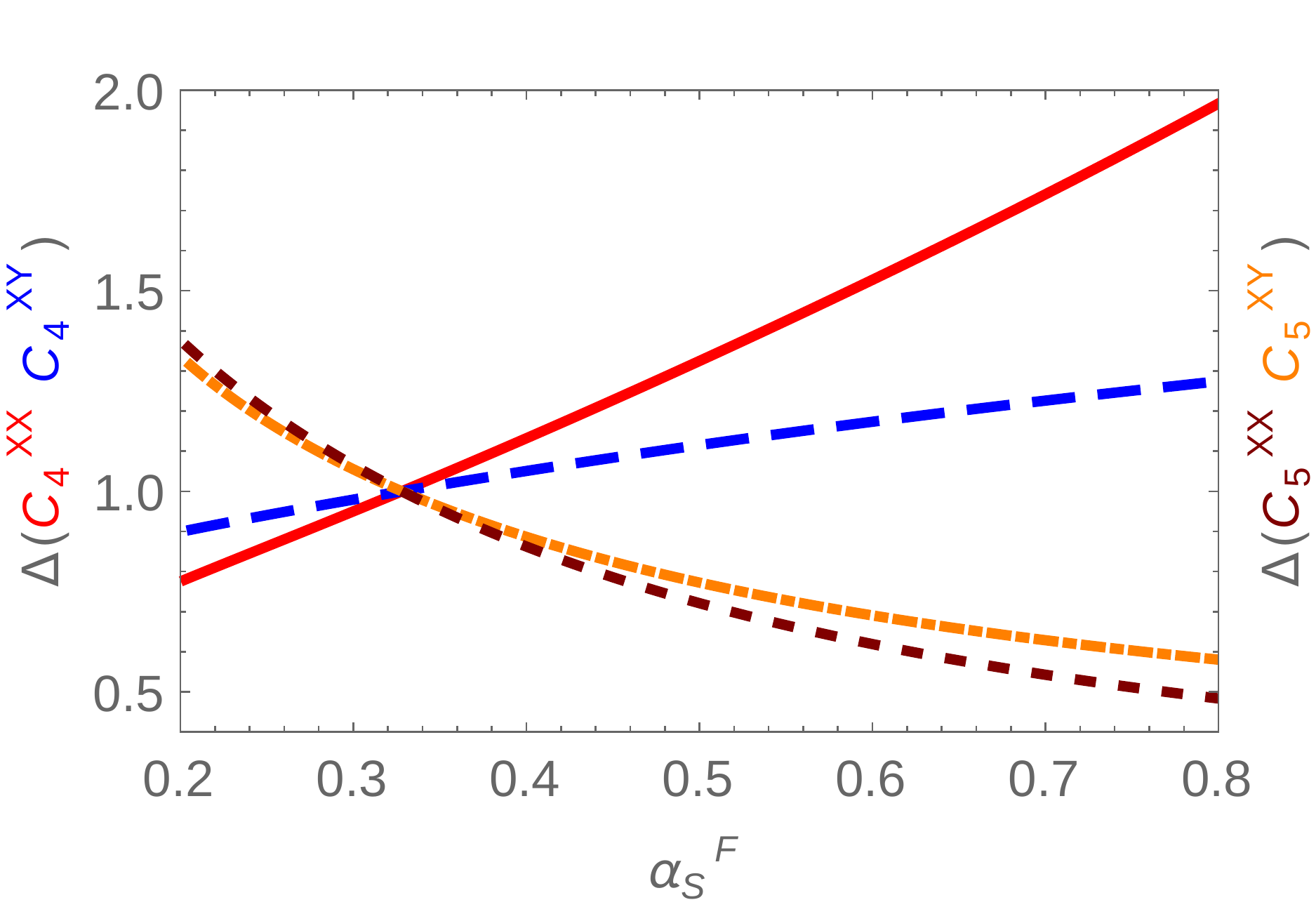}

\caption{Relative change of the limit on the short-range coefficients
  with respect to the ``frozen'' value of $\alpha_S$ at low energies.
  Here $\alpha_S^F$ represents $\tilde\alpha_S(Q^2)$ for $Q^2 \le 0.1$
  GeV$^2$.  $\Delta(C_i^{AB})$ is calculated normalizing with respect
  to the value of the coefficient derived without ``freezing'' and
  using an $\alpha_S({\rm 1 \hskip1mm GeV}^2) \simeq 0.32$.   Up
    to $\alpha_S^F \approx 0.8$, changes of the Wilson coefficients
    are roughly less than a factor of $\sim 2$.  }
\label{fig:short}
\end{figure}

\begin{figure}[t]
\includegraphics[width=1\linewidth]{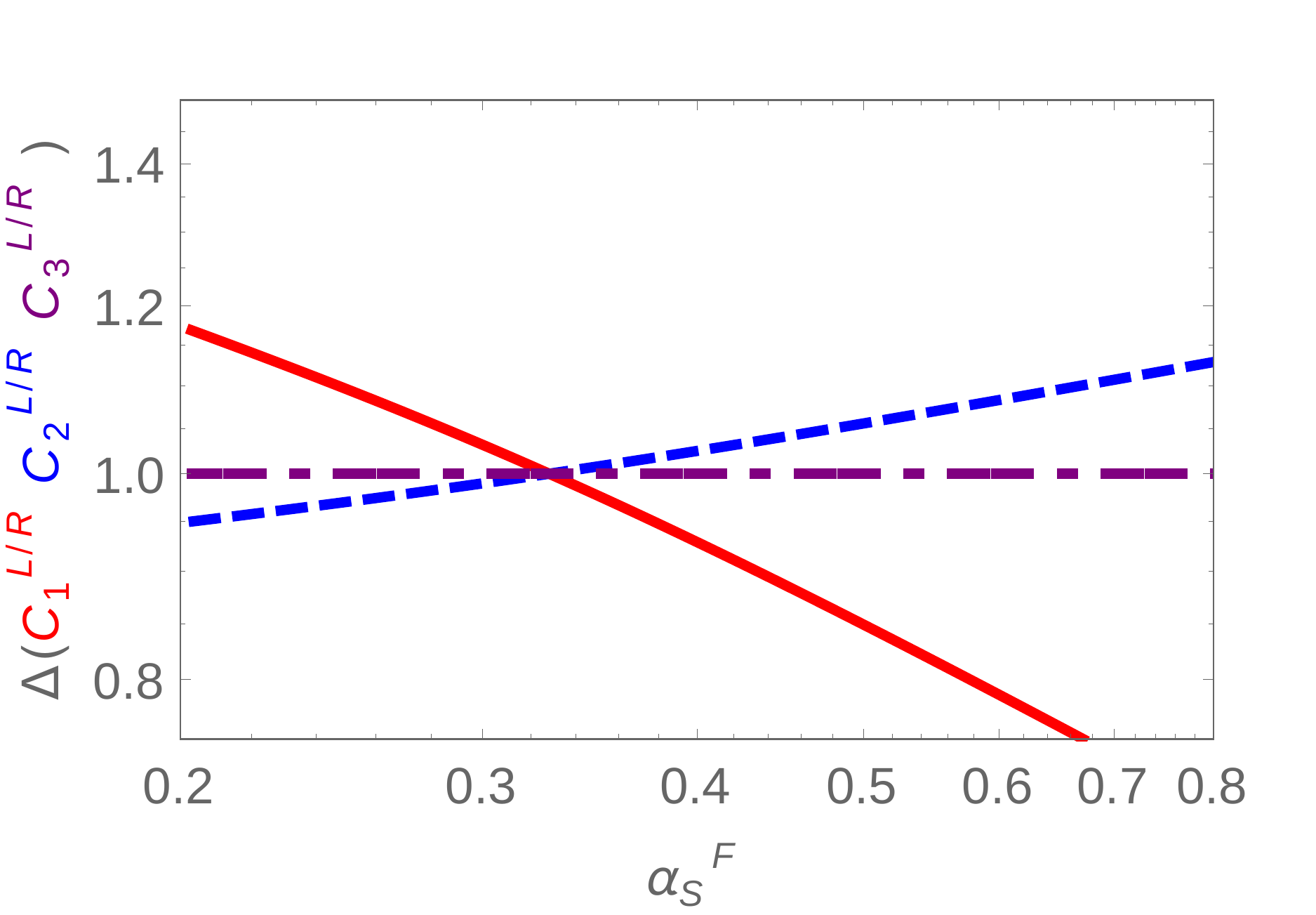}
\caption{Relative change of the limit on the long-range coefficients
  with respect to the ``frozen'' value of $\alpha_S$ at low energies.
Note the change of scale  with respect to Fig.~\ref{fig:short}.}
\label{fig:long}
\end{figure}

Our main results are then shown in Fig.~\ref{fig:short} for the 9
SRM coefficients and in Fig.~\ref{fig:long} for the 6 LRM
coefficients. Both figures show the change of the WCs with respect to
their numerical values calculated without freezing, i.e.
$\Delta(C_i^{AB})=C_i^{AB}(\alpha_S^F)/C_i^{AB}(\alpha_S({\rm
  1\hskip1mm GeV}))$, as a function of the numerical value of
$\alpha_S^F$. Note that there are only 3 lines in
Fig.~\ref{fig:long}, because the relative changes of $C_i^L$ and
$C_i^R$ are the same. Note also, that the coefficients are calculated
for the particular case of $^{136}$Xe, but the plots for other
isotopes are very similar (since we plot $\Delta(C_i^{AB})$).

 As Fig.~\ref{fig:short} shows, all 9 short-range coefficients 
change only moderately, when $\alpha_S^F$ is varied in the window
(0.2-0.8). The two $C_3$ are the most stable coefficients, while 
 the other $C_i$ change within (less than) a factor of roughly
two. We would like to point out that it is usually argued in the
$\znbb$ decay community, that nuclear matrix elements (NME) have 
uncertainties of a factor of roughly 2 as well, thus we call the
change of the coefficients under variation of $\alpha_S^F$
``moderate'', since it is of comparable size.

\begin{figure}[t]
\includegraphics[width=1.0\linewidth]{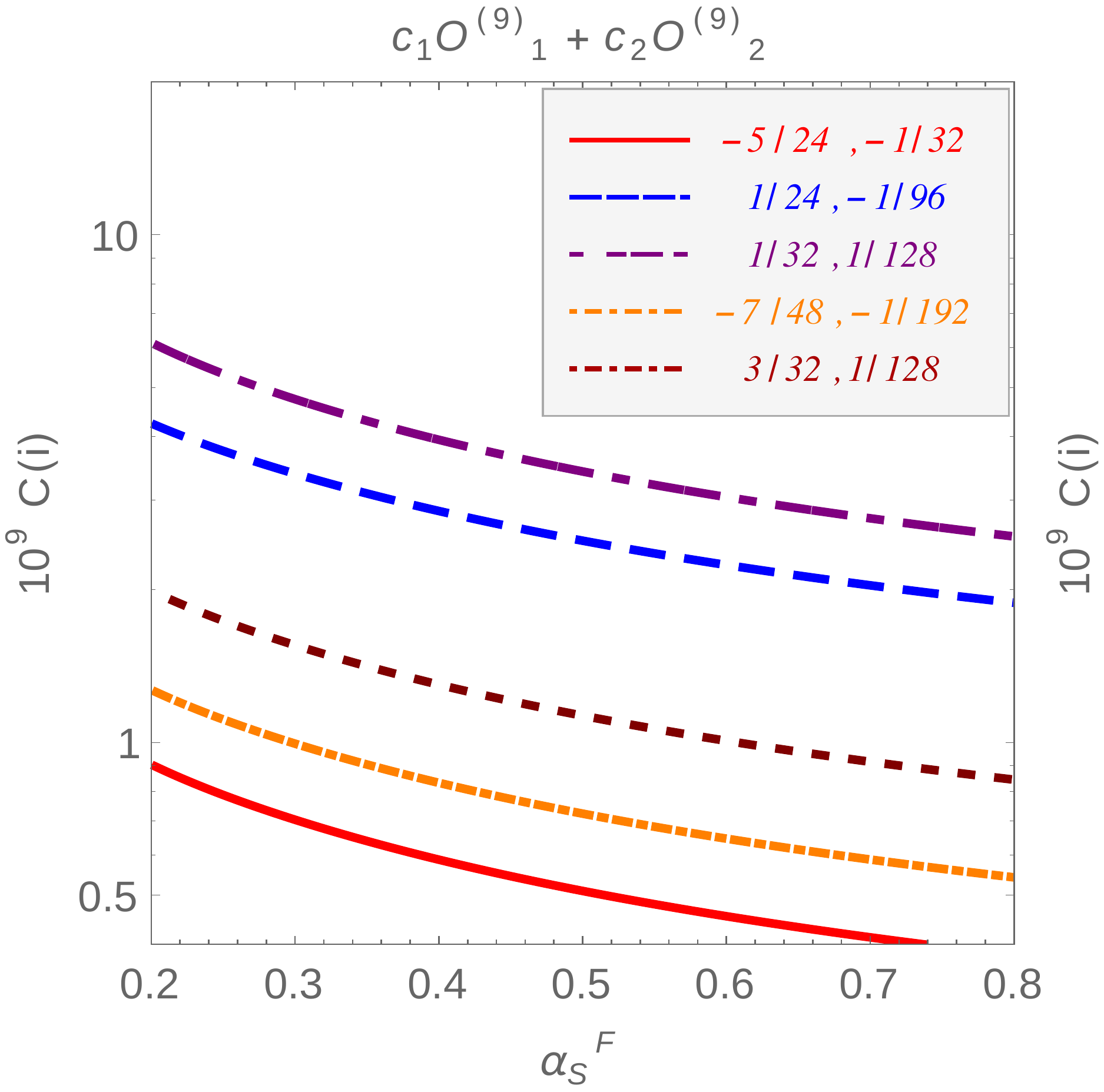}
\caption{Limits on the five linearly independent short-range
  combinations of $c_1 O_1^{(9)} + c_2 O_2^{(9)}$ that appear in the
  tree-level decomposition of the $\znbb$ decay operator, see
  \cite{Bonnet:2012kh}. For a discussion see text.}
\label{fig:comb}
\end{figure}

  Nevertheless, we would like to mention that for values of
  $\alpha_S^F \gsim 1$ larger changes of the Wilson coefficients
  result. Thus, the stability of our extrapolation rests on
  whether or not the idea of a finite, frozen value of $\alpha_S$
  is indeed correct.

 We would also like to discuss that among the operators in
Eq.~(\ref{eq:OperBasis-2}), forming the low-energy operator basis, the
tensor operators $\mathcal{O}_{2}^{(9)}$ are special in the sense that
they can never appear alone, i.e. without $\mathcal{O}_{1}^{(9)}$, in
the low-energy limit of any renormalizable high-scale model
(HSM). This is because renormalizable models do not contain
fundamental tensor interactions. Then, at low energies the effective
operators $\mathcal{O}_{2}^{(9)}$ can arise only from a Fierz
transformation of the scalar operator $\mathcal{O}_{1}^{(9)}$ in the
Lorentz and color indices. As was shown in \cite{Bonnet:2012kh} there
are 5 possible linear combinations of these two operators, which can
originate from renormalizable high-scale models (for details see
Ref.~\cite{Arbelaez:2016zlt})
\begin{eqnarray}\label{eq:HSM-operators}
\mathcal{O}_{K}^{HSM}&=& c_{1} \mathcal{O}_{1}^{(9)} + c_{2} \mathcal{O}_{1}^{(9)}
\end{eqnarray} 
We show the five combinations of coefficients $c_{1}, c_{2}$ in
Fig.~\ref{fig:comb} together with the corresponding limits on these
combinations, again as a function of $\alpha_S^F$. The plot
demonstrates that these limits are stable  (within again roughly
  a factor 2) with respect to variations of $\alpha_{S}^{F}$ for all
5 cases.

Fig.~\ref{fig:long}, on the other hand, shows that the LRM
coefficients are much more stable under variations of $\alpha_S^F$.
Note the change in the scale of the figure relative to
Fig.~\ref{fig:short}. Thus, one can conclude that both, running and
freezing, do not play an important role for the long-range part of the
$\znbb$ decay amplitude.

\section{Discussion and conclusions}
\mbox{}\\[-5mm]
We have calculated QCD corrections to $\znbb$ for both cases, the SRM
and LRM, with particular emphasis on the IR behaviour of the QCD
running coupling. We discussed the regulated form of the QCD running
coupling, using Background Perturbation Theory, which introduces the
background mass parameter $M_B$. In this treatment, the resulting
$\tilde\alpha_S(Q^2)$ ``freezes'' at values of $Q^2$ smaller than
$Q^2\le 0.1$ GeV. However, since $M_B$ in this setup is both, a
model-dependent as well as a process-dependent parameter, the exact
value of $\alpha_S^F$ is not predicted. We, therefore, showed our
results as a function of $\alpha_S^F$. We conclude that the
Wilson coefficients depend only moderately on the exact value of
$\alpha_S^F$ and it seems we can extract reliable limits on these
coefficients from $\znbb$.  We noted that in renormalizable
ultra-violet completions (``high scale models'') the tensor operator
is never generated alone, ie. it always appears in combination with
$O_1^{(9)}$.  We demonstrated that all combinations of $c_1 O_1^{(9)}
+ c_2 O_2^{(9)}$ that appear in the (tree-level) decomposition of the
$\znbb$ decay operator are also stable in the above quoted sense.

On the other hand, as a word of caution we need to mention
that the idea of a frozen $\alpha_S$ at low energy is not at present
universally accepted, see the detailed discussion in \cite{Deur:2016tte}.
Our conclusions will remain valid if $\alpha_S^F$ can be shown to lie
definitely below, say, \mbox{$\alpha_S^F \lsim (0.7-0.8)$}. For values of
$\alpha_S^F$ much larger than this, our simple-minded treatment can 
not be considered to be valid.

Finally, we would like to stress again that the present
  study does not pretend to present a well theoretically
  grounded analysis of the non-perturbative QCD effects in the
  transition region from the quark to the nucleon level description of
  $\znbb$-decay. The problem of the matching, at some low-energy
  scale, of the perturbative QCD results with the effective low-energy
  theory in terms of the nucleon fields remains open and is waiting
  for its clarification. Future work in this direction would be
  very valuable.

\bigskip

\centerline{\bf Acknowledgements}
%
This work was supported by the Spanish MICINN grants FPA2014-58183-P,
FPA2017-85216-P and SEV-2014-0398, and PROMETEOII/2014/084
(Generalitat Valenciana), and by Fondecyt (Chile) under grants,
No. 1150792 and No. 3160642 as well as \mbox{CONICYT} (Chile) Ring ACT
1406 and Basal FB0821.


%

\end{document}